# KWT-Tiny: RISC-V Accelerated, Embedded Keyword Spotting Transformer

Aness Al-Qawlaq, Ajay Kumar M, Deepu John
University College Dublin, Ireland

*Abstract*—This paper explores the adaptation of Transformer-based models for edge devices through the quantisation and hardware acceleration of the ARM Keyword Transformer (KWT) model on a RISC-V platform. The model was targeted to run on 64kB RAM in bare-metal C using a custom-developed edge AI library. KWT-1 was retrained to be 369 times smaller, with only a 10% loss in accuracy through reducing output classes from 35 to 2. The retraining and quantisation reduced model size from 2.42 MB to 1.65 kB. The integration of custom RISC-V instructions that accelerated GELU and SoftMax operations enabled a 5x speedup and thus ~5x power reduction in inference, with inference clock cycle counts decreasing from 26 million to 5.5 million clock cycles while incurring a small area overhead of approximately 29%. The results demonstrate a viable method for porting and accelerating Transformer-based models in low-power IoT devices.

*Keywords—Transformer models, IoT, custom hardware, GELU RISC-V, SoftMax, hardware acceleration, quantization, KWT*

## I. INTRODUCTION

Since its conception in the famous landmark paper "Attention is All You Need" by Google in 2017 [6], the Transformer model has met and exceeded state-of-the-art performance in various applications; chief amongst which is natural language processing (NLP). When compared to convolutional neural networks (CNNs) and recurrent neural networks (RNNs) in the NLP field, the Transformer has exhibited better handling of long-range dependencies, superior semantic feature extraction abilities, as well as the ability to be very effectively parallelized [1]. This has directly led to the onset of popular large language models such as ChatGPT, which are transformer-based models with hundreds of billions of trainable parameters. The success of the Transformer model at NLP tasks led researchers to investigate the use of the Transformer models for applications that have traditionally been performed by the CNN architecture, such as Computer Vision [3], Keyword Spotting [4], and Gesture Detection [5].

However, this comes at the expense of a large model size and increased computational cost. Therefore, many Transformer models are beyond the computational capabilities of resource-constrained hardware. This presents a challenge for the growing Internet of Things (IoT) movement which aims to enable cheap, resource-constrained devices to invoke cutting-edge models in a power-efficient manner. This problem is twofold – edge AI systems struggle with both the large inference-time memory requirements, as well as long execution times associated with performing the required operations. In order to solve this issue, current literature focuses on quantising the transformer model, as well as introducing custom hardware to accelerate inference.

There have been several different hardware acceleration approaches in literature. Some, like AI-RISC [13] and AIfES [14] target lower-power applications, but do not target the Transformer architecture specifically. Other approaches, such as FlexACC [15] and DaDianNao [2] focus on designing neural network processors which support many different machine learning architectures, outperforming general-purpose processors in speed and energy consumption. There have also been approaches that target the specific acceleration of the Transformer architecture. RISC-VTF [1] proposes acceleration through custom RISC-V instructions, and $A^3$ [7] proposes acceleration of the attention mechanism through approximation.

However, while current literature addresses the issue associated with compute time, they are typically addressed through the lens of high-power, high-performance systems without much focus on the memory-related challenges associated with pushing Transformers to be invoked on the edge. Additionally, they tend to emulate Transformer operations to measure the speedup as opposed to accelerating a real model. This makes it challenging to gauge the accuracy degradation due to hardware acceleration.

On the other hand, this paper involves a more holistic approach. A state-of-the-art keyword spotting transformer model is trained, modified, and quantised on a PC. Then, its inference operations are implemented in bare-metal C and compiled for a RISC-V platform. Lastly, the RISC-V platform is supplemented with custom hardware blocks, and the processor's instruction set is extended with a few custom instructions to accelerate the inference of the model.

This paper investigates the performance degradation of Transformers as they are downsized to fit on low-power devices. It also focuses on the general framework involved in training and accelerating Transformer-based models in the context of embedded systems, with ARM's Keyword Transformer (KWT) [4] as the key example. Additionally, the custom hardware and instructions are designed to be flexible, accelerating any Transformer-based model, including encoder-decoders, vision transformers, and BERT.

## II. BACKGROUND: THE KEYWORD TRANSFORMER (KWT)

The Transformer is a novel machine learning architecture first proposed by Google in 2017 for NLP purposes, where Google first introduced the concept of self-attention [6]. Up until the creation of the Transformer, deep learning models for NLP uses were based on CNNs or RNNs [8]. These architectures, while powerful, relied on fixed-length

This work was supported in part by 1) Science Foundation Ireland under the US-Ireland R&D Program 2) Science Foundation Ireland through the SFI Centre for Research Training in Machine Learning (18/CRT/6183) and Microelectronic Circuits Centre Ireland
*Code available at: https://github.com/anessk01/kwt_tiny

representations of input sequences. CNNs and RNNs struggle with long-term dependencies due to the maximum path length between inputs and outputs being large. The shorter the paths between inputs and outputs in a model, the easier it is to learn long-range dependencies [9], improving performance. The attention mechanisms within the Transformer achieve this goal by reducing forward and backward signal path length.

The earlier mentioned KWT model is built on the Vision Transformer (ViT) architecture, which is a post-norm, encoder-only Transformer specifically built for computer vision tasks. It has exhibited 98.6% accuracy on the Google Speech Commands (GSC) dataset. As shown in Fig 1, a single inference pass of KWT starts with a conversion of the raw input audio signal to a Mel-scale spectrogram $(X)$ with time windows $t = 1, ..., T$ and frequencies $f = 1, ..., F$, also known as Mel-Frequency Cepstral Coefficients (MFCC). The next steps involve splitting the spectrogram $X \in \mathbb{R}^{T \times F}$ into separate flattened time domain patches, on which a linear projection is applied. This linear projection maps the spectrogram to a higher dimension $d$ using a projection matrix $W_0 \in \mathbb{R}^{F \times d}$. The positional embeddings $X_{pos} \in \mathbb{R}^{(T+1) \times d}$ are then applied. The resulting patches are fed into the Transformer's $l$'th block $(X_l)$.

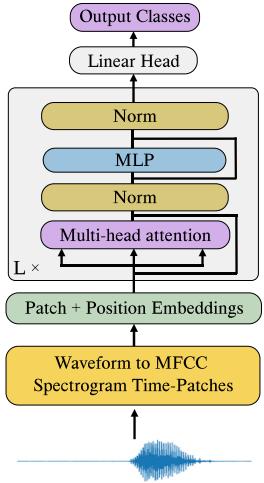

Fig. 1. KWT inference pass, figure modified from [1].

The signal $X_l$ is split into three vectors: Query (Q), Key (K), and Value (V). Next, the dot product of Q and K is found and normalised by $\sqrt{d_h}$ where $d_h$ is the dimensionality of each attention head. Next, the SoftMax (2) operation is applied to the result, and then multiplied by the value vector. The result of the multiplication is the self-attention (SA) vector (1):

$$SA(X_l) = softmax\left(\frac{QK^T}{\sqrt{d_h}}\right)V \quad (1)$$

Where:

$$softmax(\vec{x})_i = \frac{e^{x_i}}{\sum_{j=1}^{K} e^{x_j}} \quad (2)$$

And weights $W_Q$, $W_K$ and $W_V$ are obtained through training:

$$Q = X_l W_Q, \quad K = X_l W_K, \quad V = X_l W_V \quad (3)$$

The results from (1) with mean $(\mu)$ and variance $(\sigma^2)$ are then normalised as in (4):

$$\widehat{X_l} = \frac{\vec{X_l} - \mu}{\sqrt{\sigma^2}} \quad (4)$$

After this step, pre-computed scale $(\vec{\gamma})$ and shift $(\vec{\beta})$ are applied to each element $\widehat{X_l}$ to produce the final normalised result $y_i$ (5):

$$y_i = \gamma_i \cdot \widehat{X_l} + \beta_i \quad (5)$$

The normalised result is then passed through a multilayer perceptron (MLP) block. MLPs are a type of feedforward neural network with behaviour as defined in (6):

$$FFN(x) = \text{GELU}(xW_1 + b_1)W_2 + b_2 \quad (6)$$

Where, as in [10], $erf$ is the Gauss error function:

$$GELU(x) = x \times \frac{1}{2}\left(1 + \text{erf}\left(\frac{x}{\sqrt{2}}\right)\right) \quad (7)$$

The output from the MLP is normalised as in (5) and goes through a final linear mapping before the output class is produced as in (8).

$$Linear(x) = xW_1 + b_1 \quad (8)$$

As can be seen, this process involves many matrix multiplications, SoftMax operations, and GELU calculations. Additionally, it can also be seen that intermediate results between layers must be freed from memory whenever no longer needed since it is very limited on IoT platforms.

III. KWT-TINY

The parameters of the KWT model which achieved the 96.9% accuracy figure are shown in Table I. As is evident, such a large number of parameters, each of which is a 32-bit floating point number, could not be loaded onto the few kB of RAM available in most low-power embedded systems. The embedded system being used in this paper is the lowRISC Ibex [17], which has the specifications listed in Table II.

TABLE I. KWT-1 MODEL SPECIFICATIONS [1]

| Attribute | Specification |
| --- | --- |
| # Parameters | 607k |
| Output Classes | 35 |
| Accuracy | 96.9% |

Due to the limited memory resources on this platform, a much smaller KWT-based model, KWT-tiny, had to be trained. Through an iterative approach, the layers with the least impact on inference accuracy were removed. These were found to be the depth layers, which are the sequential transformer layers where the output of one is fed into the input of the next. One transformer layer was found sufficient enough to support two output classes, as opposed to the 12 layers in the original KWT for 35 output classes.

TABLE II.   lowRISC Ibex Specifications

| Attribute | Specification |
|---|---|
| RAM | 64 kB |
| Clock Speed | 50 MHz |
| FPU | Not Available |

To further satisfy the memory constraints without sacrificing too much accuracy, the input MFCC was down-sampled from the original [40, 98] to [16, 26], which proved to be a reasonable balance between memory constraints and accuracy constraints. More changes in dimensions of each of the layers were made to further ensure that the floating-point weights of the model, as well as program logic, were able to fit in the available 64kB of RAM. KWT-Tiny was trained using the Torch-KWT [11] library. These changes are summarised in Table III.

TABLE III.   KWT-Tiny vs KWT-1

| Attribute | Significance | KWT-1 | KWT-Tiny |
|---|---|---|---|
| INPUT_DIM | Dimensions of Input Spectrogram | [40, 98] | [16, 26] |
| PATCH_DIM | Dimensions of a Single Spectrogram Patch | [40, 1] | [16, 1] |
| DIM | Size of layer normalisation vector | 64 | 12 |
| DEPTH | Number of transformer layers in series | 12 | 1 |
| HEADS | Number of parallel attention computation blocks | 1 | 1 |
| MLP_DIM | Size of MLP network | 256 | 24 |
| DIM_HEAD | Size of each attention head | 64 | 8 |
| SEQLEN | Size of attention scores matrix | 99 | 27 |
| OUTPUT CLASSES | Number of unique word-classes that can be discerned | 35 | 2 |

As is evident in Table III, KWT-Tiny can only discern two output classes, meaning that it is ideal for the detection of a single keyword such as "Hey Google" or "Alexa", as opposed to KWT-1 which can discern between 35 different output classes in the GSC dataset.

The accuracy of KWT-tiny was tested on the full Google Speech Commands dataset, where the two output classes were "dog" and "notdog". The results of the accuracy of KWT-tiny as well as its parameter count when compared to KWT-1 are shown in Table IV.

As can be seen in Table IV, a 369x reduction in model size of KWT-1 resulted in a small ~10% reduction in model accuracy. This was largely due to the success of the iterative approach which correctly identified the least important layers and downsized them accordingly. This finding indicates that Transformer-based models can indeed retain high accuracy when scaled down for embedded applications if the correct layers are targeted. For KWT-Tiny, it was found that downsizing the MLP network as well as the depth of the transformer block presented the best accuracy-size trade-off, whereas overly downsizing the normalization vector led to steep accuracy loss.

TABLE IV.   KWT-Tiny vs KWT-1: Accuracy

| Attribute | KWT-1 | KWT-Tiny | % Change |
|---|---|---|---|
| # Parameters | 607k | 1646 | −99.73% |
| Memory use (Floating Point) | 2.42 MB | 6.58 kB | −99.73% |
| Accuracy | 96.9% | 87.2% | −9.7% |

IV. Quantising KWT-Tiny

Quantising the KWT-Tiny model is important for two main reasons: (1) storing model weights in INT8 as opposed to FLOAT32 reduces model size by 4 times, and (2) performing mathematical operations using floating point emulation on embedded devices with no FPU is very expensive, especially when it comes to floating point division.

KWT-Tiny-Q and its input MFCC were quantised through post-training static quantisation, whereby all model weights were multiplied by a static scale factor ($2^y$) and then quantised to INT8 as in (9).

$$W_{int} = floor(W_{float} \times 2^y) \qquad (9)$$

The scale factor was chosen to be a power of 2 to make quantisation and dequantisation as cheap as possible on an embedded platform by using bit shifts.

Intermediate residuals which would be the result of the integral multiplications of various INT8 weights are sized as INT16 to prevent too much data loss during inference. As per other transformer quantisation techniques, SoftMax and layer normalisation continued to be computed in a floating-point manner, as quantising those operations was found to be quite taxing on accuracy [12]. This would mean that the intermediate results are dequantised before being passed to the SoftMax and LayerNorm layers, the outputs of which would be re-quantised for the next layer and so on.

To decide the value of $2^y$, each audio input in the GSC dataset was fed into KWT-Tiny and also into different versions of KWT-Tiny-Q, each with different values of $2^y$ in (9). The accuracy of each version was then compared as in Table V. A scale factor too large would result in overflows in the intermediate results, and a scale factor value too small would reduce the precision of the quantised weights.

As is evident in Table V, the ideal approach involved a different scale factor for weights and inputs. This is because weights range from 0 to 1, whereas the MFCC input vector can contain elements with magnitude of a few hundred. This means that the weights can be scaled by a larger factor than the input while increasing precision and not contributing to overflows. KWT-Tiny-Q exhibited 25% of the memory consumption of KWT-Tiny with 82.5% accuracy, a loss of ~5%.

TABLE V.  KWT-TINY-Q ACCURACIES

| Scale Factor $2^y$ for Weights | Scale Factor $2^y$ for Input | Model Size | Accuracy |
|---|---|---|---|
| 8 | 8 | | 60.3% |
| 16 | 16 | | 71% |
| 32 | 32 | 1.646kB | 77.3% |
| **64** | **32** | | **82.5%** |
| 64 | 64 | | 65.2% |

## V. KWT-TINY IN BARE-METAL C

Current edge AI approaches for transferring models onto resource-constrained platforms involve the use of libraries such as TensorFlow Lite (TFLite), ONNX, or TVM. The KWT-1 model is available for usage on mobile phones with TFLite [4]. A challenge with these approaches is that the libraries themselves can be as large as a few MB. While this is no issue with modern smartphones with multiple GB of RAM, many embedded platforms would not be able to fit such libraries in memory. Another challenge with these approaches is that the embedded developer and computer architect are both abstracted away from the lower-level details of the operation of the model, which makes operation acceleration and memory optimisation for a particular platform very difficult.

To counteract these challenges, this paper proposes a simple C library for transformer-based computations. This approach maximises customisability and memory optimisation. The library is created to allow for both floating point-based computations for non-quantised models as well as INT16 operations for quantised models. The library should be compatible with most Transformer-based models. The downside with this approach is that the embedded engineer must be familiar with how to create the inference pipeline based on the provided library functions, but this should be a simpler task than developing the entire tensor library. To enable the computations in Fig. 1 and Fig. 2, the library described in Table VI was created.

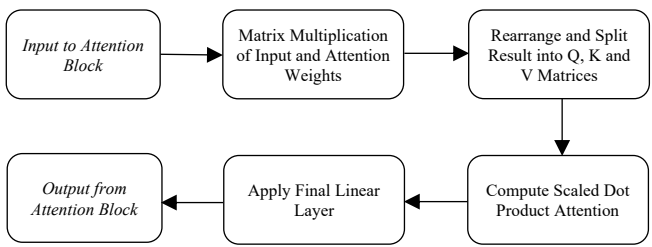

Fig. 2.  Self-Attention computation pipeline

The inference C code comprises two main sections: initialization, involving copying model hyperparameters and loading weight pointers; and the inference pipeline. However, fitting model parameters in memory alone doesn't guarantee sufficient runtime memory for calculations due to intermediate result storage. Efficient memory usage is crucial, with stack size calculated for maximal runtime needs. The linker must be configured to allocate stack memory accordingly. In this case, 60kB program memory and 4kB stack were allocated. Compiler size optimization using the "-Os" flag was necessary to fit KWT-Tiny on the Ibex platform.

TABLE VI.  C TRANSFORMER-BASED TENSOR LIBRARY

| Method | Purpose |
|---|---|
| computeMeanAndVariance() | Used in layer normalisation operations as discussed in (4). Computes mean and variance of input vector. |
| layerNorm() | Normalises every element in an input vector, applying normalisation $\beta$ and $\gamma$ as discussed in (5). |
| matrixMultiply() | Applies matrix multiplication $C = Ab$ through the use of the basic $O(n^3)$ algorithm. |
| Softmax() | Computes Softmax operation using built-in function expf() and floating point division as discussed in (2) |
| gelu() | Computes Gelu operation using built-in functions erf(), sqrt() and floating point division as in (7) |
| linear() | Computes (8) using matrixMultiply() |
| splitIntoQKV() | Splits flattened input array to query (Q), key (K), and value (V) vectors as discussed in (3). |
| scaledDotProductAttention() | Uses softmax() and matrixMultiply() to compute the scaled dot product attention result as discussed in (1) |

Additionally, memory occupied by intermediate results no longer required for the next layers' input need to be cleared. In a bare metal system with no OS support, the use of the popular memory allocation function malloc() is typically not supported. To counteract this, a manual implementation of malloc() was devised, whereby two separate arrays of pre-defined size are allocated to act as global memory banks. Their sizes are found through dry-running the pipeline and ensuring that the maximal intermediate result's size fits within one of the banks. Two banks are required as there are cases in which two residual results are required simultaneously such that they can be added together as per the pipeline in Fig. 1. In the implementation for KWT-Tiny on C, the banks were of size $SEQLEN \times MLP\_DIM$ and $SEQLEN \times DIM\_HEAD \times 3$ respectively, with attributes as defined in Table III.

## VI. HARDWARE ACCELERATION

RISC-V [16] is an open-source instruction set architecture (ISA) that is modular and extensible. It allows computer architects to extend its functionality through the addition of custom instructions which can be used to invoke custom hardware. RISC-V reserves space as per its standards for custom instructions [16]. RISC-V also comes with a mature software toolchain [19] including compilers, linkers, and assemblers. The processor in use, the lowRISC Ibex, is an open-source [17] parametrizable single-core CPU written in System Verilog and suited for low-power applications. It complies with the RV32IMC standard, which is the 32-bit version of RISC-V that supports basic integer arithmetic, fixed point multiplication, division, and a compressed instruction set extension which provides 16-bit compact instructions. It is synthesised on the Xilinx Arty-A735T FPGA. The acceleration of KWT-Tiny on a RISC-V platform can be obtained through the introduction of a few custom instructions that the processor can use to compute certain inference-time operations. Once these are introduced, they can be invoked directly using assembly language invocation through the 'asm' keyword as per the RISC-V GNU toolchain standards [19]. This approach means that the compiler toolchain does not need to be modified to detect the operations in question; allowing for more universally compilable code. The non-accelerated, non-quantised KWT-Tiny model was profiled to obtain the most compute-intensive operations during an inference run. The results are shown in Fig. 3, Fig. 4, and Fig. 5.

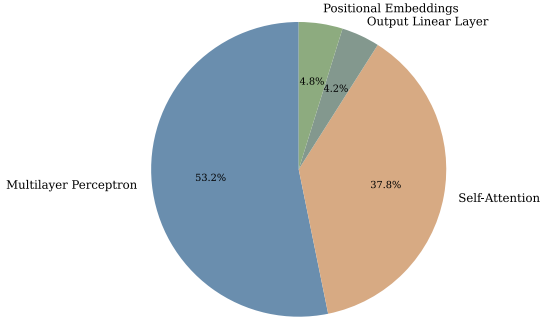

Fig. 3. Profiling of single inference run by operation

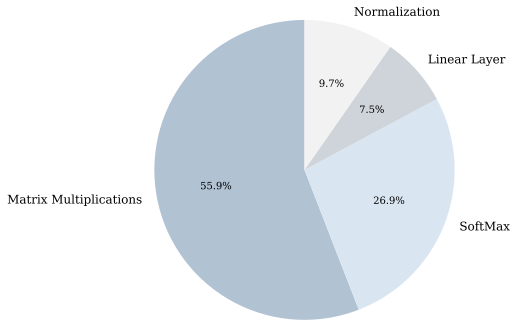

Fig. 4. Profiling of single self-attention computation by operation

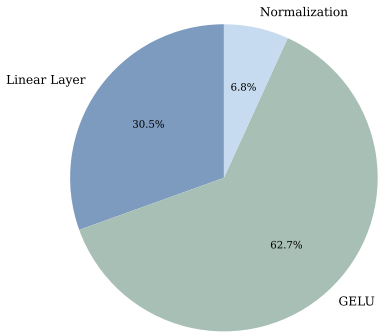

Fig. 5. Profiling of single MLP computation by operation

The GELU and SoftMax operations were found to be taxing in inference. As a result, an R-type RISC-V custom instruction (Fig. 6) is added to the Ibex decoder logic and is called directly in assembly in the C code. This custom instruction uses RISC-V's reserved "custom-1" opcode with value 7'b0101011 [16]. Once called, the decoded instruction would activate specially designed hardware blocks housed in the Ibex's modified ALU to accelerate GELU and SoftMax. GELU and Softmax operations were chosen to be accelerated as they are shared across many different deep neural network applications, not only Transformers.

| funct7 | rs2 | rs1 | funct3 | rd | opcode | R-type |

Fig. 6. R-type instruction in RISC-V

The value of funct7 remains 0 in all cases, and funct3 is used to decide the behaviour performed by the custom instruction as per Table VII.

TABLE VII. CUSTOM INSTRUCTION BEHAVIOUR

| funct3 Value | ALU Operator | Behaviour |
| --- | --- | --- |
| 3'b000 | ALU_EXP | Uses a lookup table to approximate exp(X) where X is a Q8.24 integer |
| 3'b001 | ALU_INVERT | Uses a lookup table to approximate (1/X) where X is a Q8.24 integer |
| 3'b011 | ALU_GELU | Uses a lookup table to approximate GELU(X) where X is a Q8.24 integer |
| 3'b100 | ALU_TO_FIXED | Converts a floating point number to Q8.24 representation |
| 3'b101 | ALU_TO_FLOAT | Converts a Q8.24 number to floating point representation |

To accelerate the SoftMax operation, the floating point division operation is replaced by a simple lookup table with operator ALU_INVERT. To design hardware that can compute SoftMax on a fixed-point basis, there needs to be a constrained range over which the input $x_i$ can vary. A typical way this has been avoided was through normalizing the SoftMax as follows [18], which obtains the same result as in (2):

$$softmax(\vec{x})_i = \frac{e^{\max(x)-x_i}}{\sum_{j=1}^{K} e^{\max(x)-x_j}} \quad (10)$$

This means that a lookup table can also be used to approximate the exp() through ALU_EXP. The C code would loop through a vector X, calculate ALU_EXP($X_i$) for each element, accumulate the sum, and then use ALU_INVERT(sum), then multiply the results to obtain the final SoftMax result. It was found that all values of $e^{\max(x)-x_i}$ lie between 0 and 10. It was also found that 32 divisions per unit struck a fair balance between accuracy and ROM size to be taken up. As such, the lookup tables for ALU_EXP and ALU_INV were chosen to be 320 elements of 32-bit numbers each, which is a total size of 2.56kB of ROM. The lookup tables (LUT) were indexed respectively as in (11) and (12):

$$LUT_1[z*32] \approx \frac{1}{e^z} \quad (11)$$

$$LUT_2[(z*32)-1] \approx \frac{1}{z} \quad (12)$$

As is clear in Fig. 7, GELU can be approximated very effectively through a simple piecewise approximation: if $x > 1.595$, then $GELU(x) = x$. If $x < -1.857$, then

$GELU(x) \approx 0$. If $-1.857 \leq x \leq 1.595$, then $GELU(x)$ can be approximated by a 32-element lookup table as defined in (13):

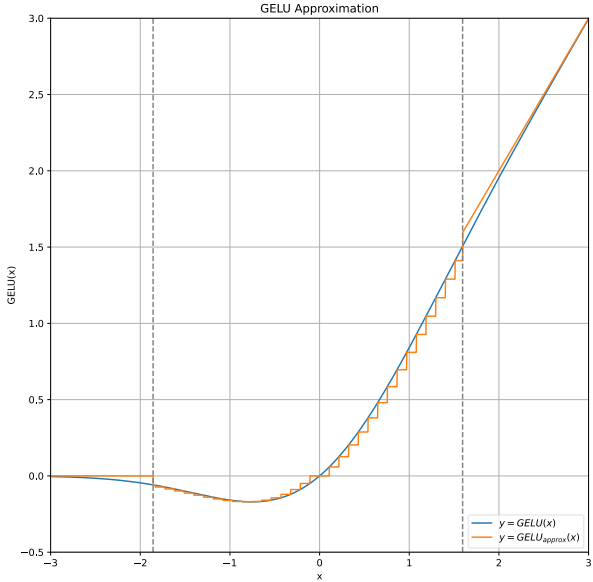

Fig. 7. Plot of $y = GELU(x)$ and $y = GELU_{approx}(x)$

$$LUT_3[x] \approx GELU(x) \quad (13)$$

The choice of the thresholds was done through a gradient descent computation that showed that this was the near-optimal choice for a 32-element LUT, with a quoted accuracy degradation of only 0.0042%. This means that the total ROM consumption is 2.69 kB for all LUTs. The results of hardware synthesis in Table VIII show a percentage increase in area of approximately 29% between baseline and modified Ibex. A comparison of all produced models is shown in Table IX.

TABLE VIII. SYNTHESIS RESULTS ON ARTY-A735T

| Attribute | Baseline Ibex | Modified Ibex | Overhead (%) |
|---|---|---|---|
| LUT | 5092 | 7368 | **10.94** |
| DSP | 10 | 16 | **6.67** |
| FF | 5276 | 6074 | **1.92** |
| BRAM | 16 | 16 | **0.00** |

TABLE IX. COMPARISON OF MODELS

| Attribute | KWT-1 | KWT-Tiny | KWT-Tiny-Q | KWT-Tiny-Q (+Hardware) |
|---|---|---|---|---|
| # Parameters | 607k | **1646** | 1646 | 1646 |
| Model Size | 2.42 MB | 6.584 kB | **1.646 kB** | 1.646 kB (+2.69 kB ROM) |
| Program Size | - | 58.8 kB | **44.4 kB** | 44.6 kB |
| Inference Clock Cycles | - | $26 \times 10^6$ | $13 \times 10^6$ | $\mathbf{5.5 \times 10^6}$ |
| Accuracy | 96.9% | **87.2%** | 82.5% | $\approx 80\%$ |

## VII. CONCLUSION

This paper investigated and accelerated Transformers from the lens of embedded systems, training, quantising, and accelerating ARM's Keyword Transformer through custom instructions on a RISC-V platform, resulting in a **5x** speedup at the cost of ~7% accuracy and 29% area overhead. The paper introduces KWT-Tiny, a two-output class keyword spotter for embedded applications that is 369 times smaller than KWT-1 at the cost of 10% accuracy. The paper also proposes a novel C Transformer library and a novel GELU acceleration technique.